\documentclass[traditabstract]{aa} 

\usepackage{graphicx}
\usepackage{txfonts}
\usepackage{url}
\usepackage{natbib}
\bibpunct{(}{)}{;}{a}{}{,} 
\usepackage[utf8]{inputenc}
\usepackage{subfigure}
\usepackage{verbatim} 
\begin{document}

   \title{Non-linear propagation of kink waves to the solar chromosphere}
   \author{M. Stangalini$^{1}$, F. Giannattasio$^{1}$, S. Jafarzadeh$^{2}$}
   \institute{$^{1}$ INAF-OAR National Institute for Astrophysics, 00040 Monte Porzio Catone (RM), Italy\\
   $^{2}$  Institute of Theoretical Astrophysics, University of Oslo, P.O. Box 1029 Blindern, N-0315 Oslo, Norway\\
   \email{marco.stangalini@inaf.it}}

  \abstract 
{Small-scale magnetic field concentrations (magnetic elements) in the quiet Sun are believed to contribute to the energy budget of the upper layers of the Sun's atmosphere, as they are observed to support a large number of MHD modes. In recent years, kink waves in magnetic elements were observed at different heights in the solar atmosphere, from the photosphere to the corona. However, the propagation of these waves has not been fully evaluated. Our aim is to investigate the propagation of kink waves in small magnetic elements in the solar atmosphere.
We analysed spectropolarimetric data of high-quality and long duration of a photospheric quiet Sun region observed near the disk center with the spectropolarimeter CRISP at the Swedish Solar Telescope (SST), and complemented by simultaneous and co-spatial broad-band chromospheric observations of the same region. 
Our findings reveal a clear upward propagation of kink waves with frequency above $~2.6$ mHz. Moreover, the signature of a non-linear propagation process is also observed.\\
By comparing photospheric to chromospheric power spectra, no signature of an energy dissipation is found at least at the atmospheric heights at which the data analysed originate. 
This implies that most of the energy carried by the kink waves (within the frequency range under study $< 17$ mHz) flows to upper layers in the Sun's atmosphere.}  

   \authorrunning{M. Stangalini}
	\titlerunning{Buffeting induced kink waves in magnetic elements}
\maketitle

\section{Introduction}
Magneto-hydrodynamic (MHD) waves  are ubiquitous in the solar atmosphere. Both large-scale magnetic structures, extending over several Mm in length, and small-scale magnetic concentrations (magnetic elements) at scales close to the resolution limit of modern solar telescopes ($\sim120-150$ km on the solar photosphere), host MHD waves of different forms \citep[e.g. kink, sausage, Alfvén;][]{1978SoPh...56....5R, 1981A&A....98..155S, Edwin1983, Roberts1983, Musielak1989, 1998ApJ...495..468S, Hasan2003, Musielak2003a, Khomenko2008, Fedun2011, 2012ApJ...755...18V, 2012A&A...538A..79N}. Since their discovery, these waves were immediately recognized as fundamental for the energy budget of the solar upper atmosphere. This hypothesis became even more attractive after the availability of very high-resolution spectropolarimetric data that have provided strong observational proofs that magnetic elements also cover a significant fraction of the solar photosphere \citep{2010ApJ...723L.164L, 2012A&A...539A...6B}. \\
The ubiquitous although non-homogeneously distributed \citep{2014A&A...561L...6S} magnetic elements are perturbed by the external photospheric plasma flows and are advected in the solar photosphere \citep{2011ApJ...743..133A, 2012ApJ...759L..17L, 2013ApJ...770L..36G, fabio04, 2014arXiv1405.0677G}.
Recent observations of the photosphere have revealed that different types of waves can coexist and interact in the same magnetic element \citep{2013A&A...554A.115S}.
\citet{2012ApJ...746..183J} reported of propagating compressible waves in magnetic elements with periods of $110-600$ s in both the photosphere and the chromosphere. Moreover, \citet{2013ApJ...768...17M} have observed torsional modes \citep{2007Sci...318.1572E} generated by vortices in the solar photosphere. These are just two examples demonstrating the large variety of waves that have been observed so far in magnetic elements.\\ 
\citet{2014ApJ...784...29M} have investigated the propagation of incompressible waves to the chromosphere, finding a good agreement between photospheric and chromospheric velocity power spectra at frequencies below $8$ mHz.\\ 
Among the many MHD waves that magnetic elements can support, kink waves induced by the solar convection have been longly considered as a viable mechanism for transferring energy from the photosphere to the upper layers of the Sun's atmosphere \citep{1997ApJ...486L.145K, Hasan2003, Musielak2003, Musielak2003a, 2008ApJ...680.1542H}.\\
Kink waves have been reported at different layers in the solar atmosphere; from the lower photosphere  \citep[e.g.][]{2011ApJ...740L..40K}, to the chromosphere \citep[e.g.][]{2013A&A...549A.116J} and the corona \citep[e.g.][]{2011Natur.475..477M}.\\
\citet{refId0} have observed the presence of subharmonics, with a fundamental period of 7.6 min, consistent with the granular time scale. This constitutes an observational evidence for the excitation of kink waves by the solar convection. The same authors also argued that the detected subharmonics might be the signature of chaotic excitation. Indeed, subharmonics represent a general characteristic of complex chaotic systems, in the presence of strong forcing \citep{2009arXiv0910.3570S, 2010arXiv1002.3363S}.\\
By using both simulations and observations,  \citet{2011ApJ...740L..40K} found that the majority of magnetic elements have horizontal velocities between 0 and 1 km/s, and $6\%$ of them have velocities in excess of 2 km/s. The same authors argued that this significant fraction of magnetic elements may contribute to heat the upper atmospheric layers, since velocities over 2 km/s can efficiently drive kink waves \citep{1993ApJ...413..811C}. More on this, \cite{2013A&A...549A.116J} have observed very short duration velocity pulses up to 15 km/s in the internetwork chromospheric magnetic elements that are well in excess of their mean value ($\sim2$ km/s), and that may significantly contribute to the energy budget of the upper layers of the Sun.\\
  \begin{figure*}
  \centering
  \includegraphics[trim=0cm 0cm 0cm 0cm, clip, width=8.cm]{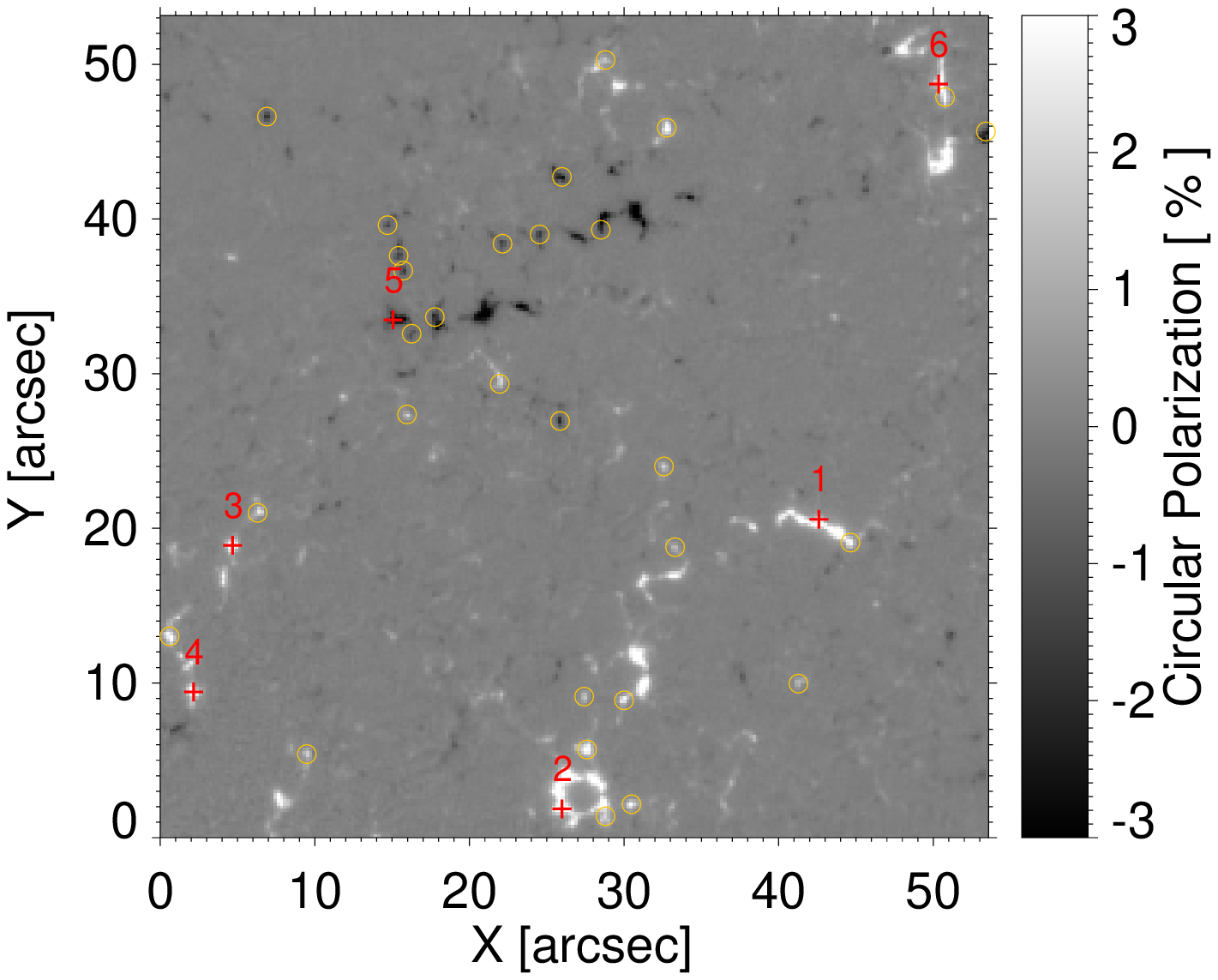}
  \includegraphics[trim=0cm 0cm 0cm 0cm, clip, width=8.cm]{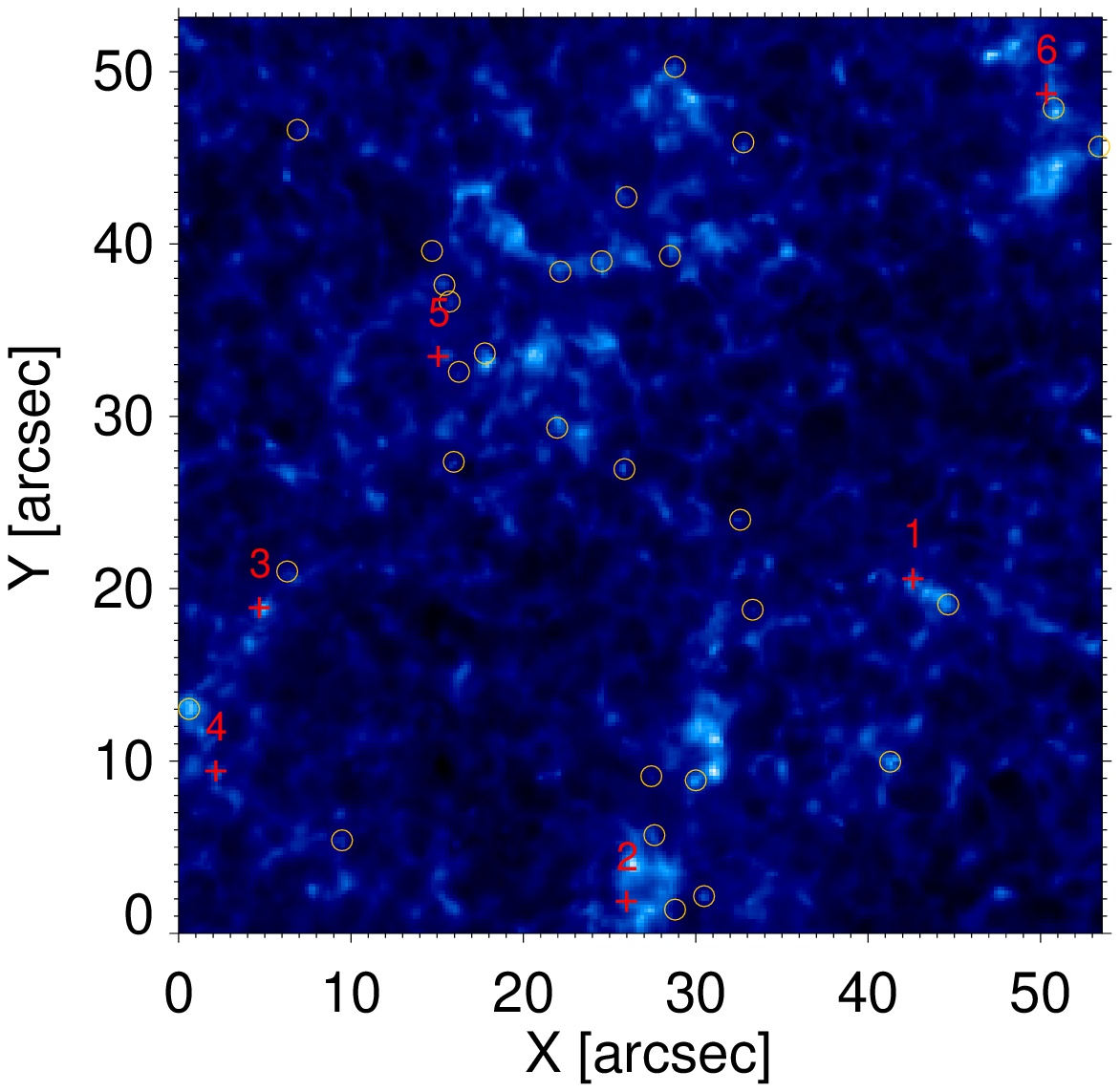}\\
  \includegraphics[trim=0cm 0cm 0cm 0cm, clip, width=8.cm]{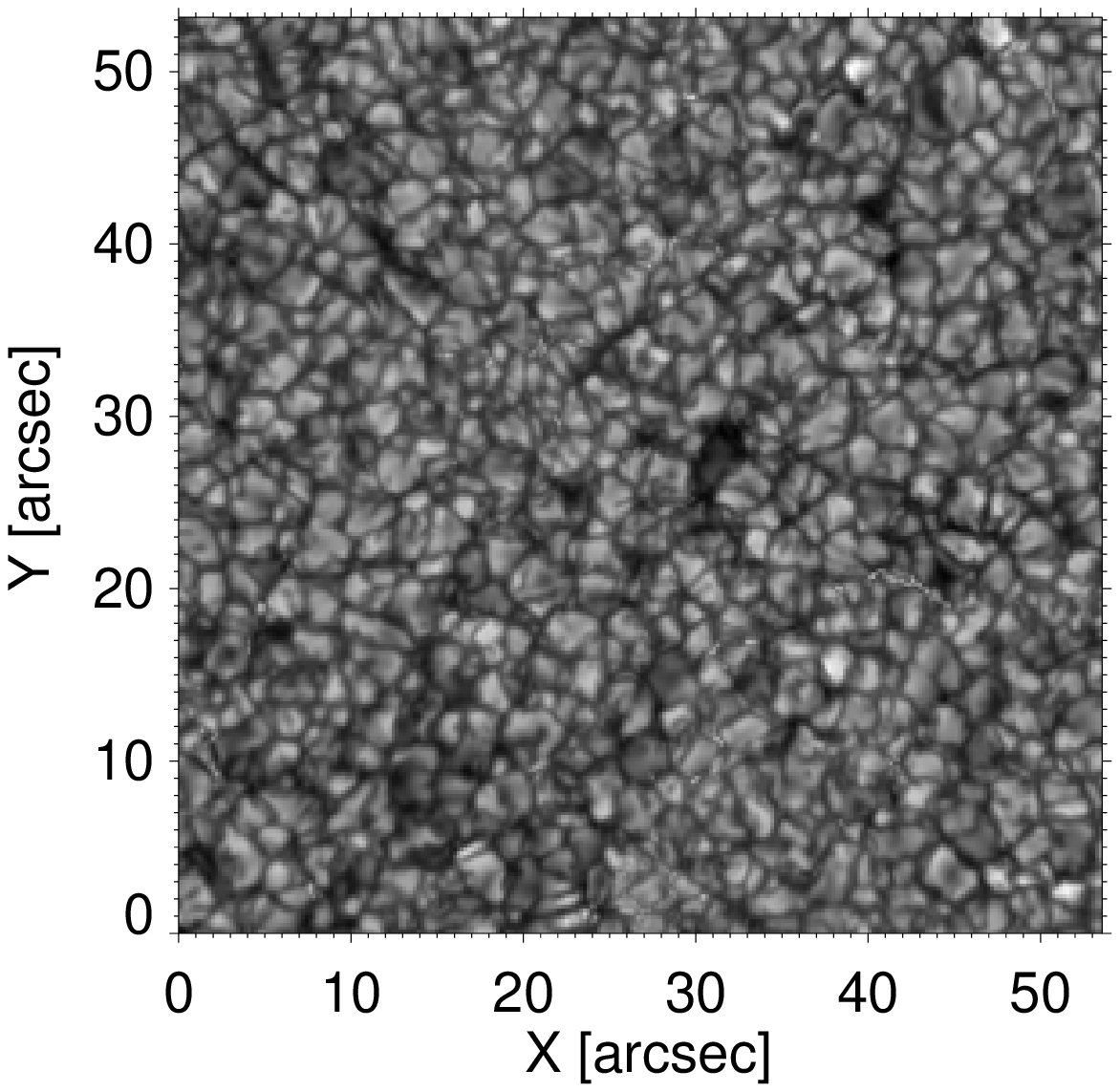}
   \caption{Example of the data analysed in this study. Left: photospheric circular polarization map saturated between $-3 \%$ and $3\%$. Center: chromospheric broad band image of the same region. The 35 longest-lived magnetic elements used in the analysis are highlighted by yellow circles. The red crosses mark the six elements whose power spectra are shown in Fig. \ref{Fig:spectra} (see text for more details). Right: continuum image taken at 395.3 nm with $1 \AA$ bandpass.}
    \label{Fig:maps}
   \end{figure*} 
In addition to the aforementioned detection of subharmonics, \citet{2013A&A...559A..88S} observed the presence of high frequency peaks of up to $10-12$ mHz in the power spectra of kink oscillations in the solar photosphere. These high frequency oscillations are well above the cutoff frequency expected for kink waves, which is always smaller than the acoustic cutoff at $\sim 5.3$ mHz \citep{1981A&A....98..155S}.
Therefore, they are expected to propagate upwards. 
Although the physical properties of the magnetic elements analyzed were quite similar, \citet{2013A&A...559A..88S} found that each individual magnetic element had its own signature in the power spectrum. This demonstrates that the kink oscillations are largely influenced by the ambient surrounding conditions.\\
\citet{2014ApJ...784...29M} have also studied the average power spectra of kink waves in both the chromosphere and the corona. Their main finding is that a considerable part of energy is dissipated between the chromosphere and the corona itself. This was proved by the observed damping of the coronal power spectra, although the chromospheric and coronal data were not sampled simultaneously, and were not co-spatial. \\ 
However, this study, as well as those reporting the presence of kink oscillations at different heights in the Sun's atmosphere, suffers lacking of information on the propagation of these waves between different atmospheric layers, due to lack of co-spatial and simultaneous data from different heights in the solar atmosphere. \\
To overcome this limitation, we analysed high-cadence and high-resolution simultaneous data of the solar photosphere and chromosphere, and investigated the propagation of kink waves in small magnetic elements in the quiet Sun, thus complementing previous results reported in the literature.

\section{Data set and methods}
The data set analysed in this work was acquired on 2011 August 6 with the imager spectropolarimeter CRISP \citep{2006A&A...447.1111S, 2008ApJ...689L..69S} at the Swedish Solar Telescope \citep[SST, ][]{2003SPIE.4853..370S}, and was aimed to the study of a quiet Sun region observed at disk center.
It consists of a full-Stokes spectropolarimetric time series of spectral scans acquired at $30$ wavelengths in the photospheric Fe I doublet at 630 nm, with a spectral sampling of $\sim$0.0044 nm.
In addition, simultaneous and co-spatial chromospheric broadband images taken at the core of the Ca II H line at $396.9$ nm were acquired.
For the spectral lines the following core-formation heights are assumed: i) $230$ km for the Fe I line at $630.1$ nm, ii) $200$ km for the Fe I line at $630.2$ nm \citep{2012A&A...548A..80F}, and iii) $450$ km for the Ca II H line at $396.9$ nm \citep{2013A&A...549A.116J}.
  \begin{figure}
  \includegraphics[width=8cm, clip]{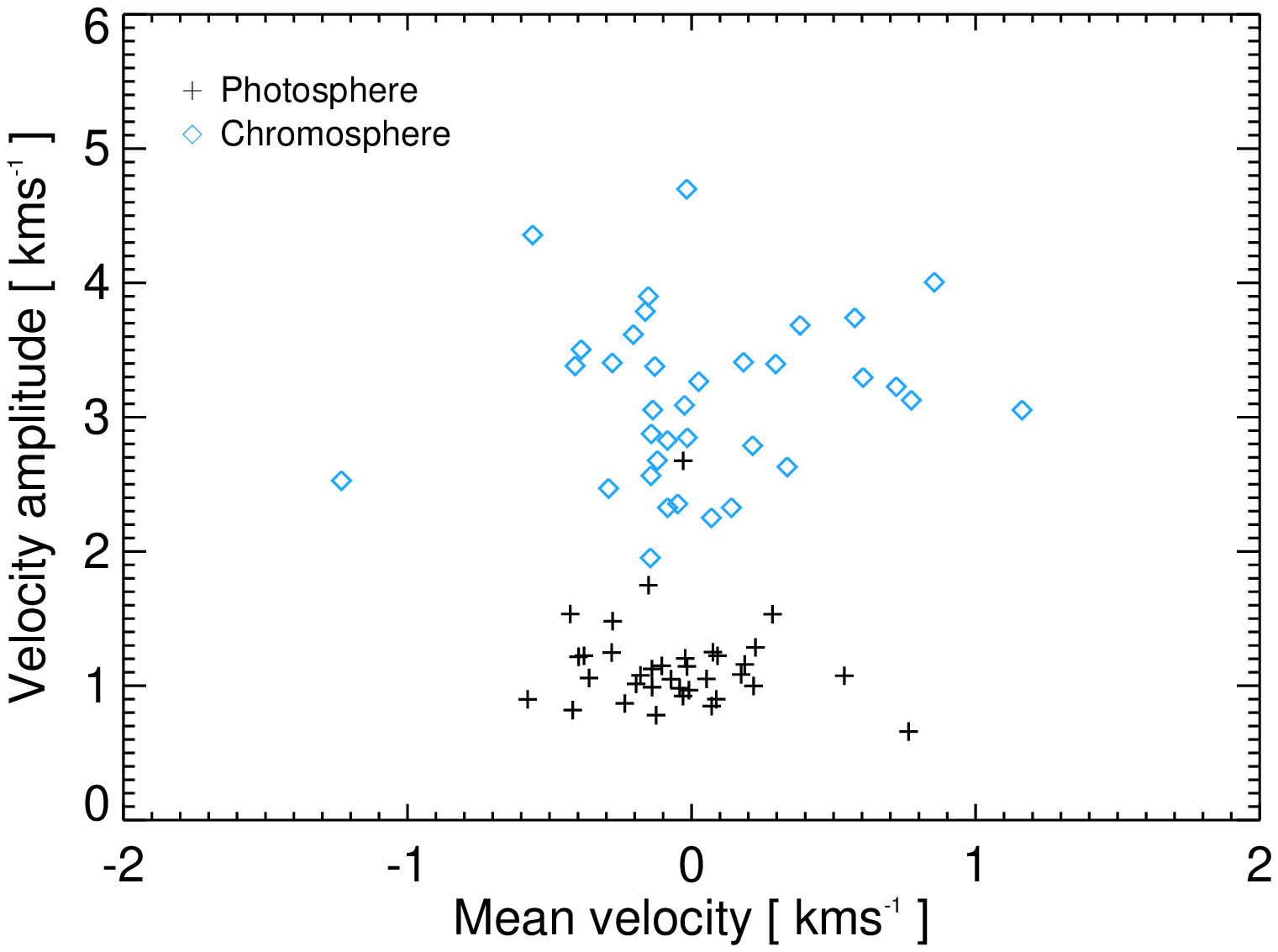}
   \caption{Chromospheric and photospheric velocity amplitude of the 35 analysed magnetic elements, as a function of their mean velocity.}
  \label{Fig:sigmavx}
  \end{figure} 
The observations started at 07:57:39 UT and lasted for $47$ minutes with a cadence of the spectral scans of $28$ s ($100$ spectral scans).
The pixel scale was $\sim0".059$/pixel for the Fe I data and $0".034$/pixel for the Ca II H data. 
The Ca II H and Fe I data were co-aligned and remapped to the same FoV of area $\sim53"\times53"$.
The theoretical full-width-half-maximum (FWHM) of the point spread function (PSF) at $630$ nm is $0".16$ or $120$ km on the solar photosphere. This constituted the diffraction limit of the telescope. The spectral resolution of CRISP at these wavelengths is 0.0055 nm \citep{2015A&A...573A..40D}.\\
The standard CRISP calibration procedure was applied to the data. 
This includes the CRISPRED calibration pipeline \citep{2015A&A...573A..40D}, and the MOMFBD \citep[Multi Object Multi Frame Blind Decomposition,][]{MSnoort05} restoration aimed at limiting seeing-induced distortions in the images. 
  \begin{figure*}
  \subfigure [Power spectra magnetic element n.1]{\includegraphics[width=6cm, clip]{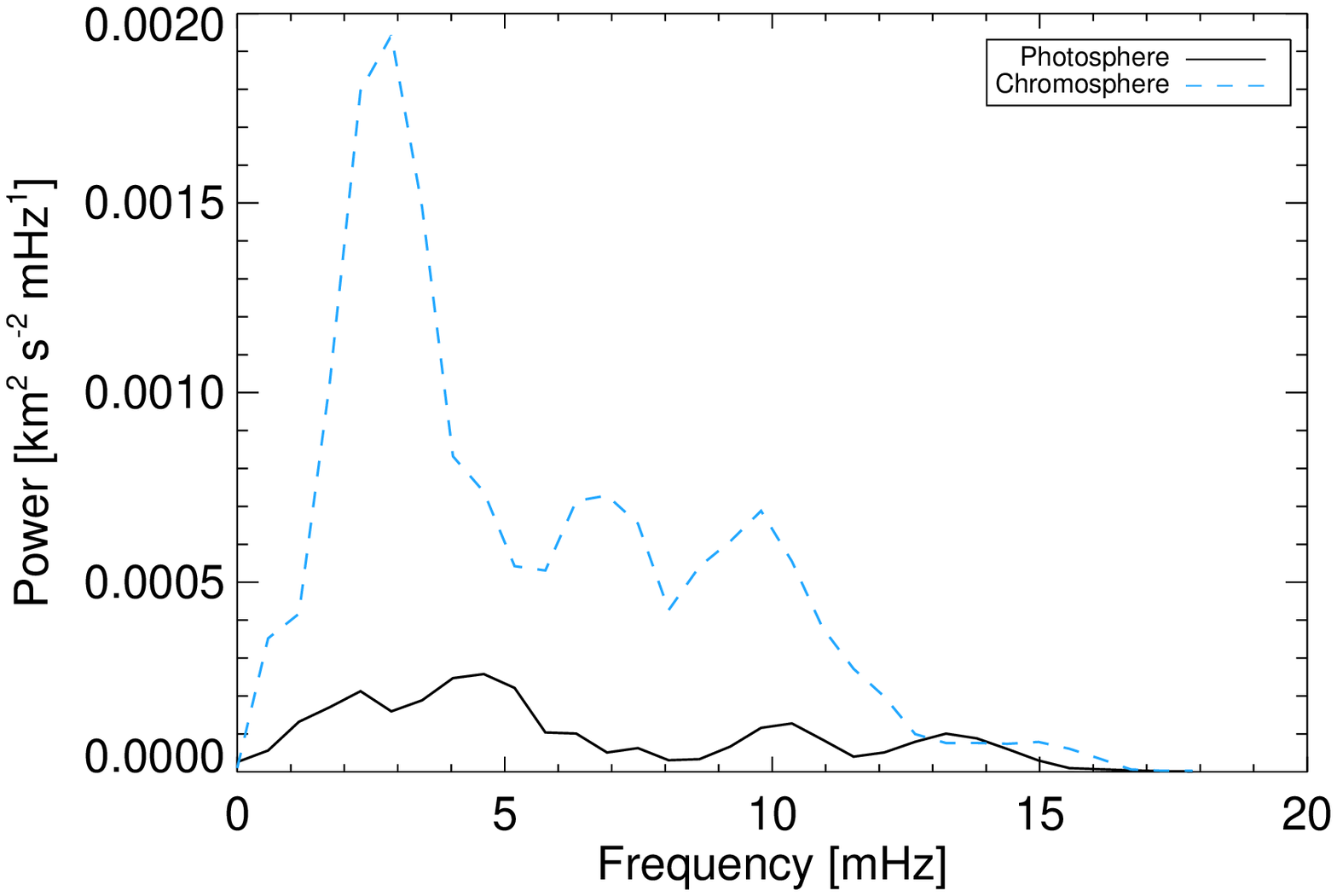}}
 \subfigure [Power spectra magnetic element n.2]{ \includegraphics[width=6cm, clip]{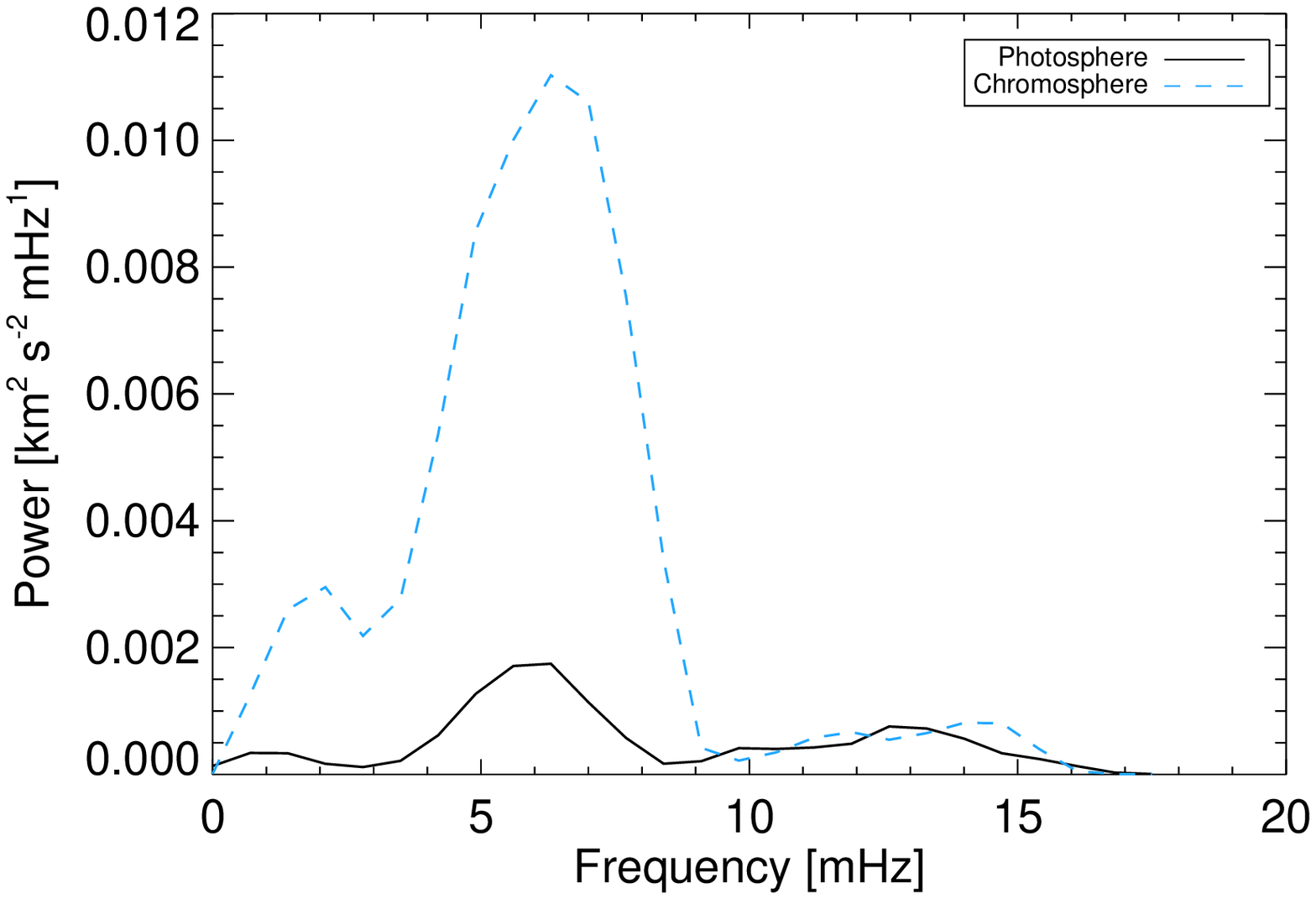}}
 \subfigure [Power spectra magnetic element n.3]{ \includegraphics[width=6cm, clip]{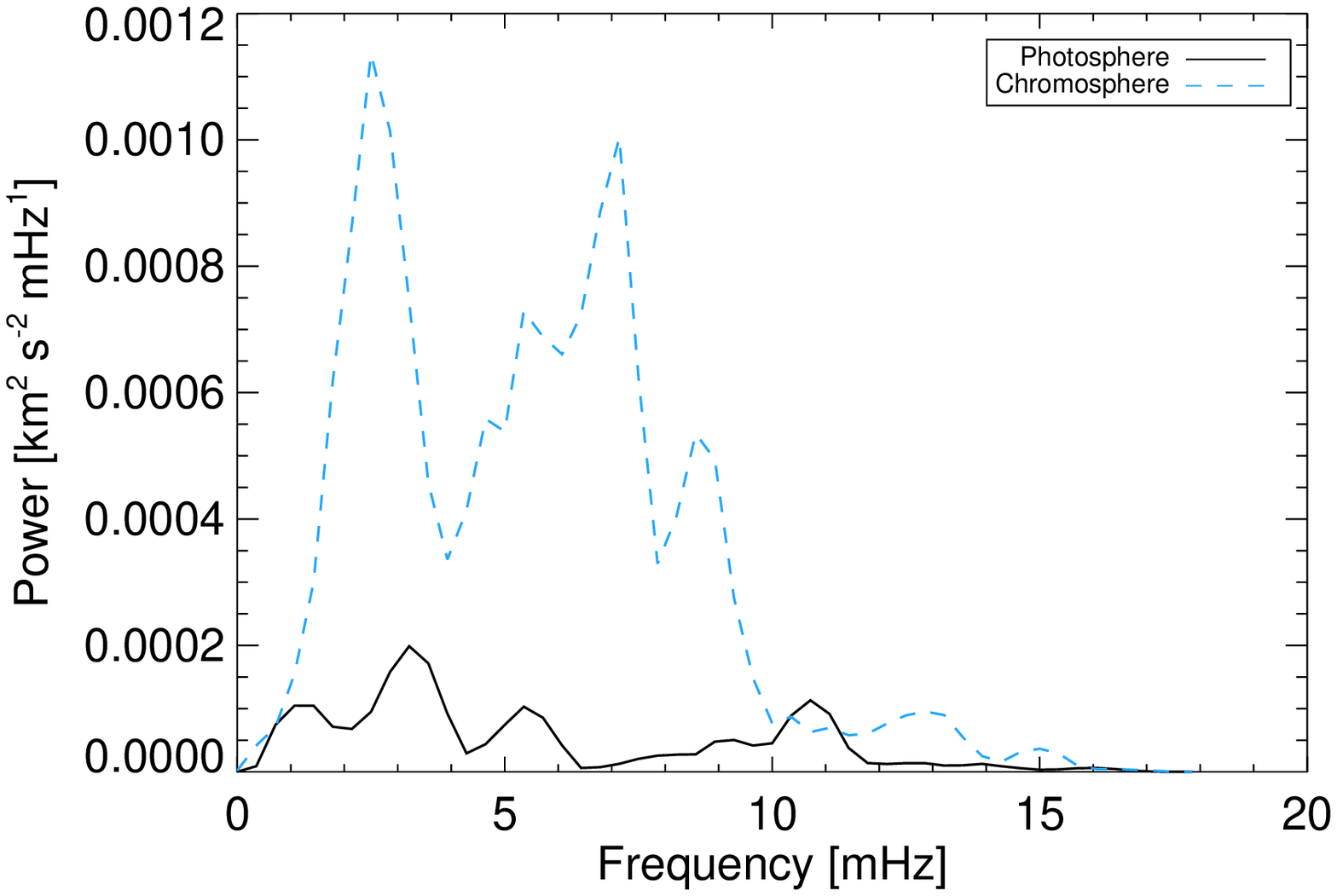}}\\
  \subfigure [Confidence level magnetic element n.1]{  \includegraphics[width=6cm, clip]{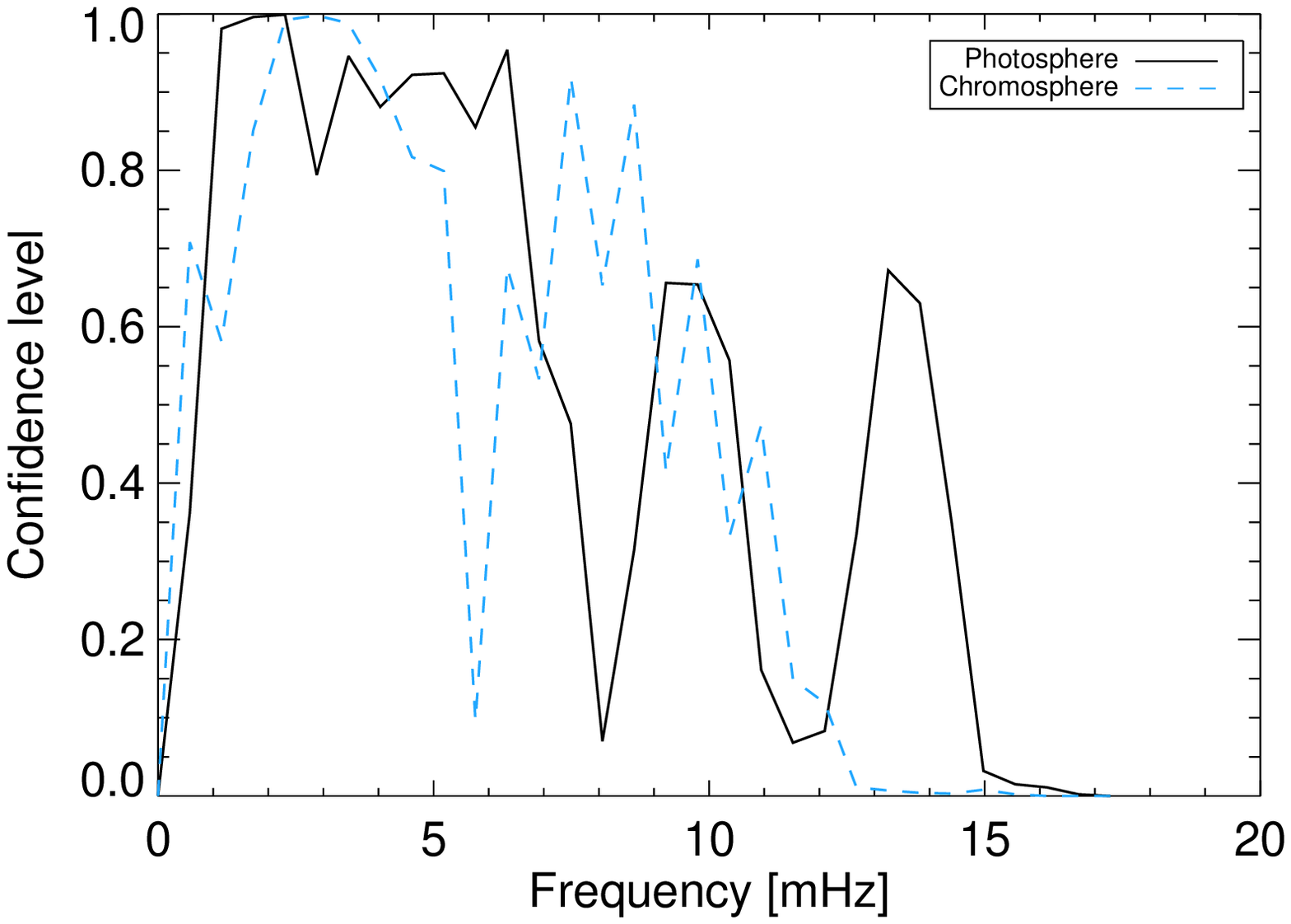}}
 \subfigure [Confidence level magnetic element n.2]{ \includegraphics[width=6cm, clip]{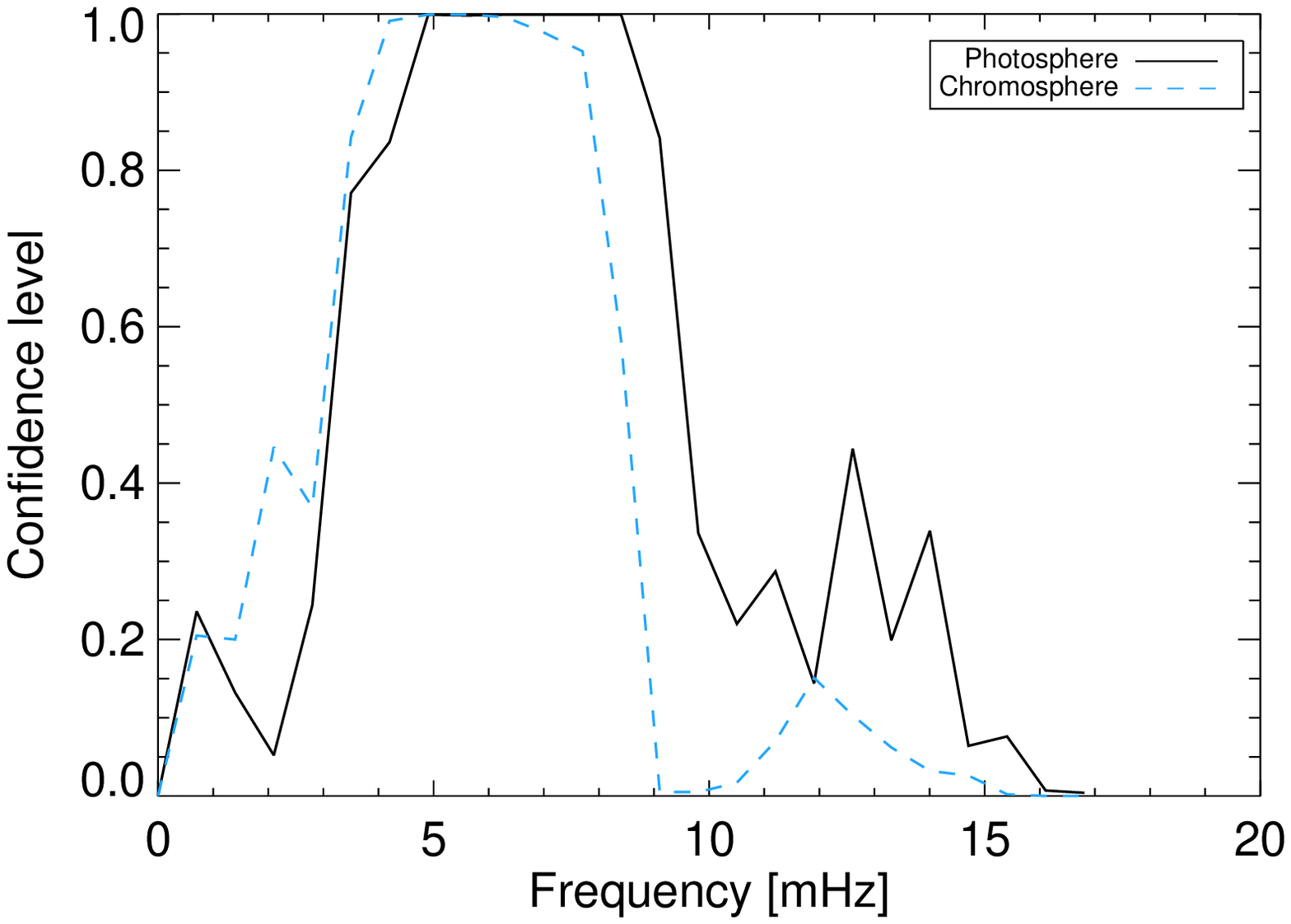}}
 \subfigure [Confidence level magnetic element n.3]{ \includegraphics[width=6cm, clip]{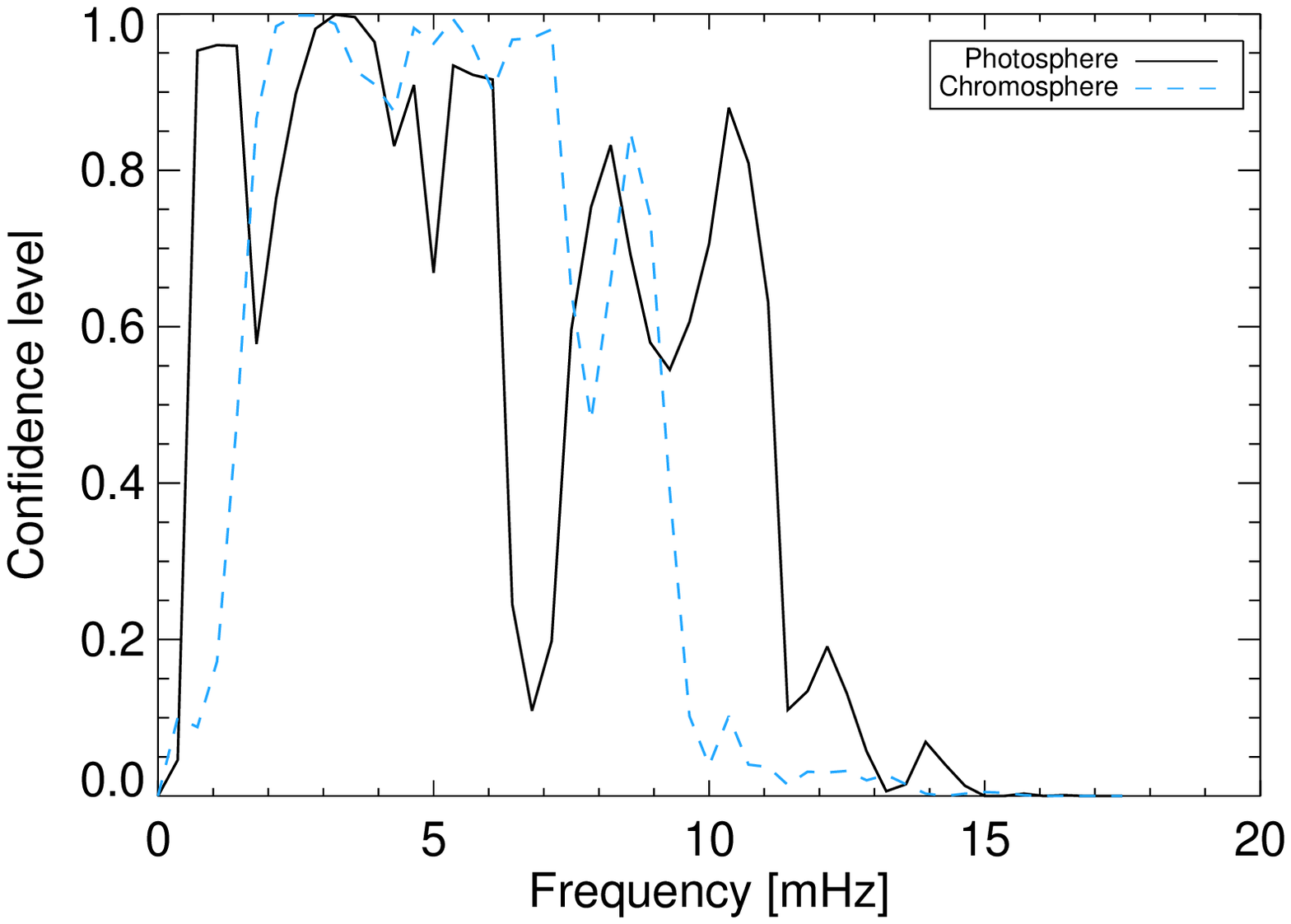}}\\
 \subfigure [Power spectra magnetic element n.4]{ \includegraphics[width=6cm, clip]{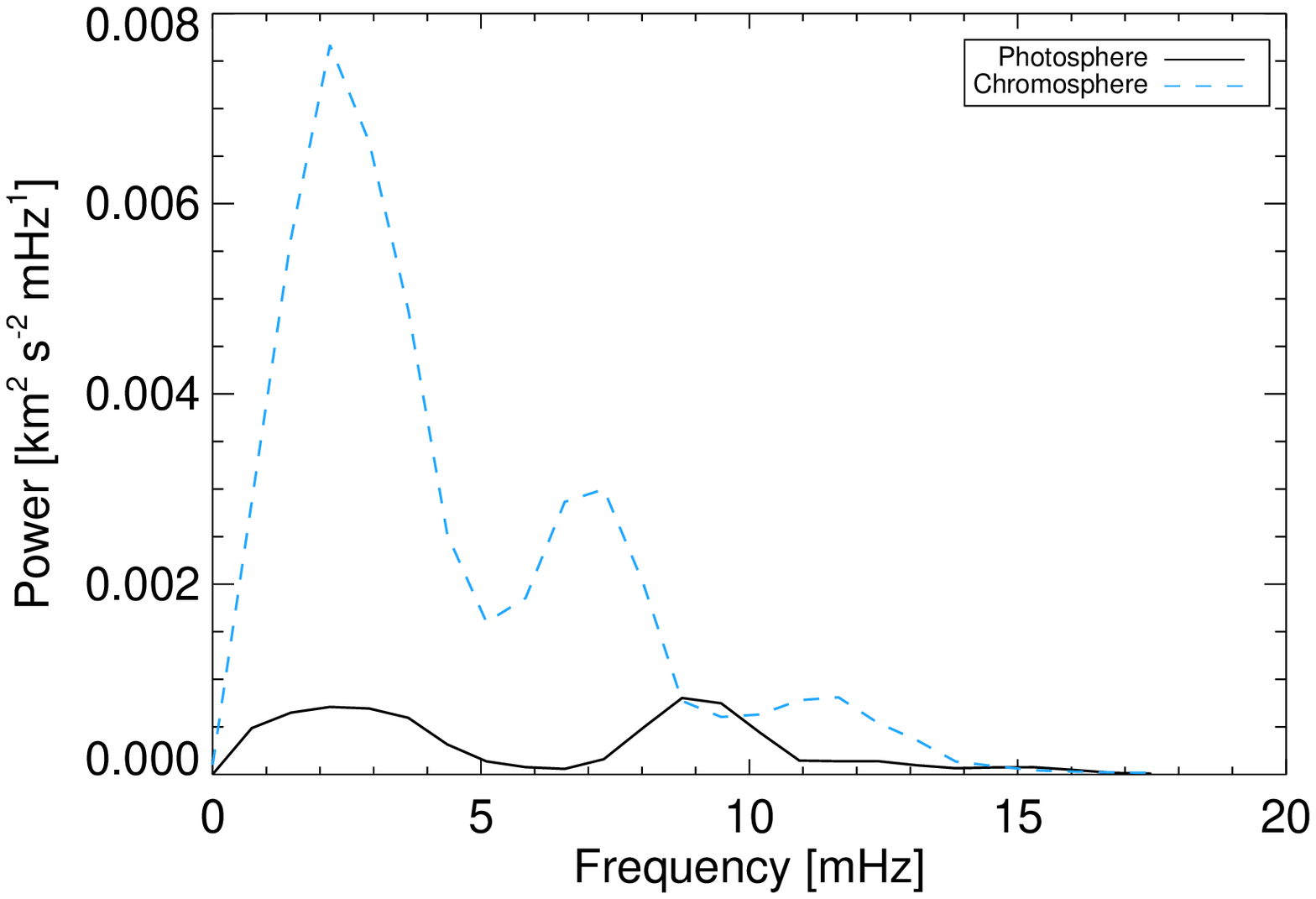}}
 \subfigure [Power spectra magnetic element n.5]{ \includegraphics[width=6cm, clip]{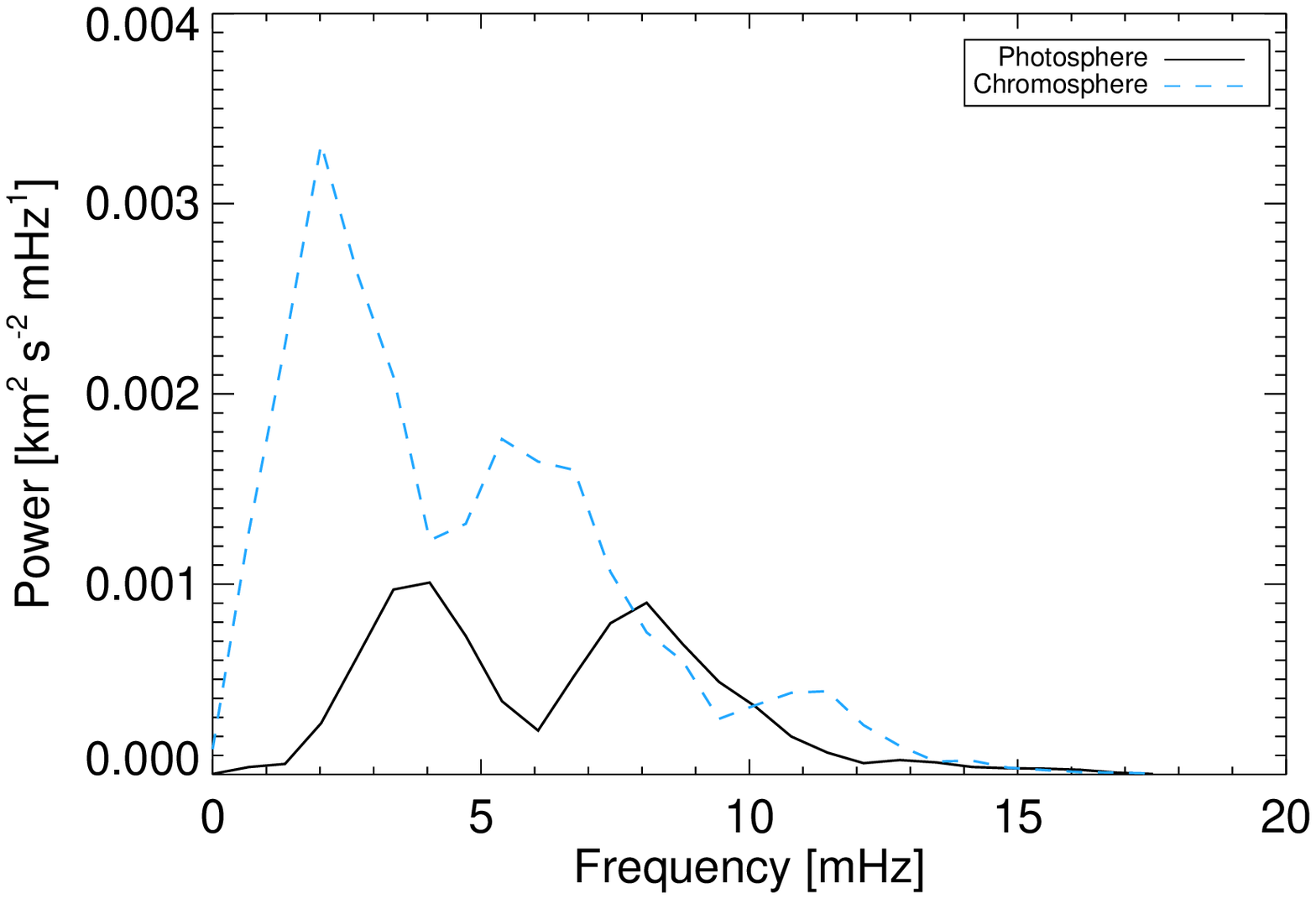}}
 \subfigure [Power spectra magnetic element n.6]{ \includegraphics[width=6cm, clip]{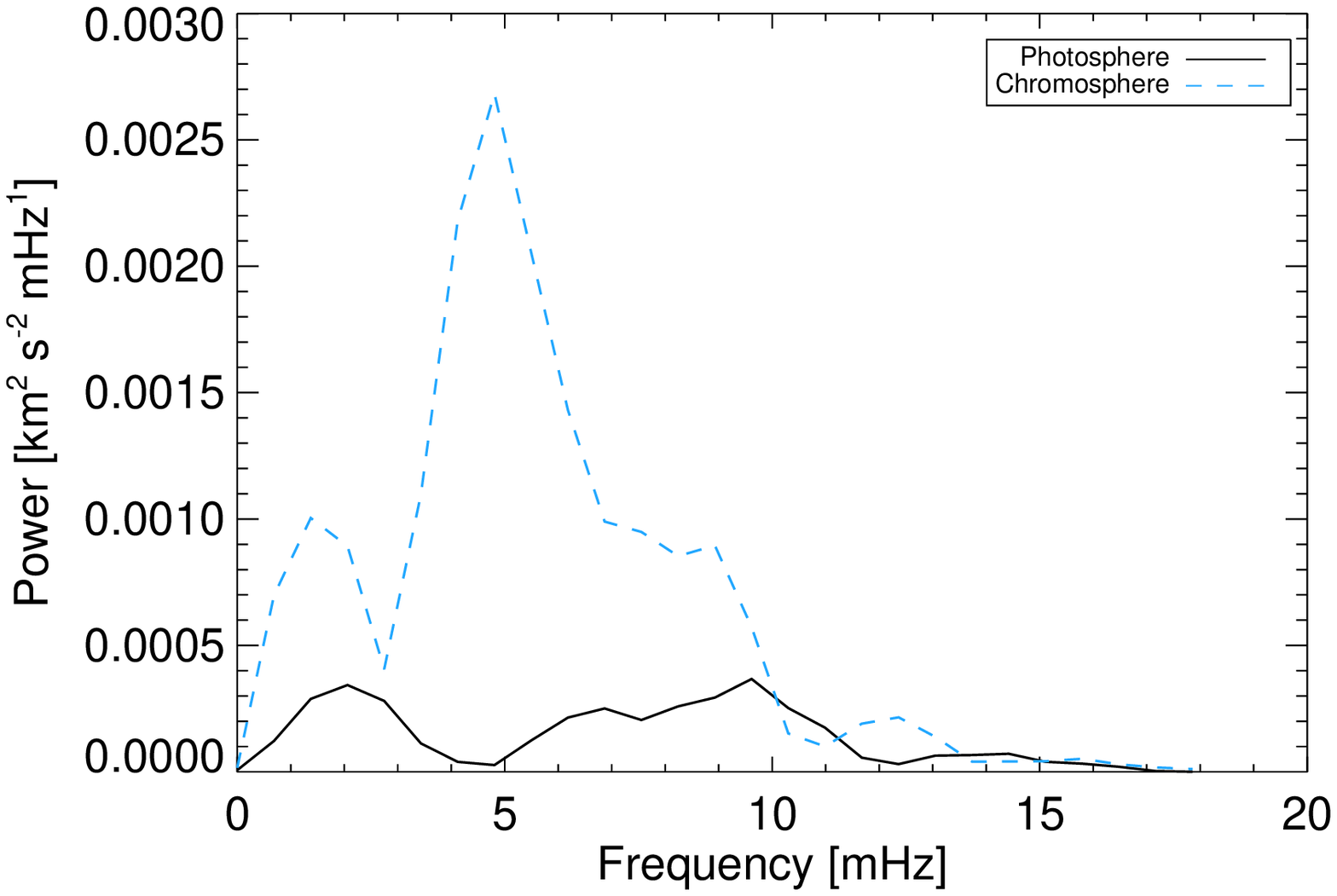}}\\
  \subfigure [Confidence level magnetic element n.4]{  \includegraphics[width=6cm, clip]{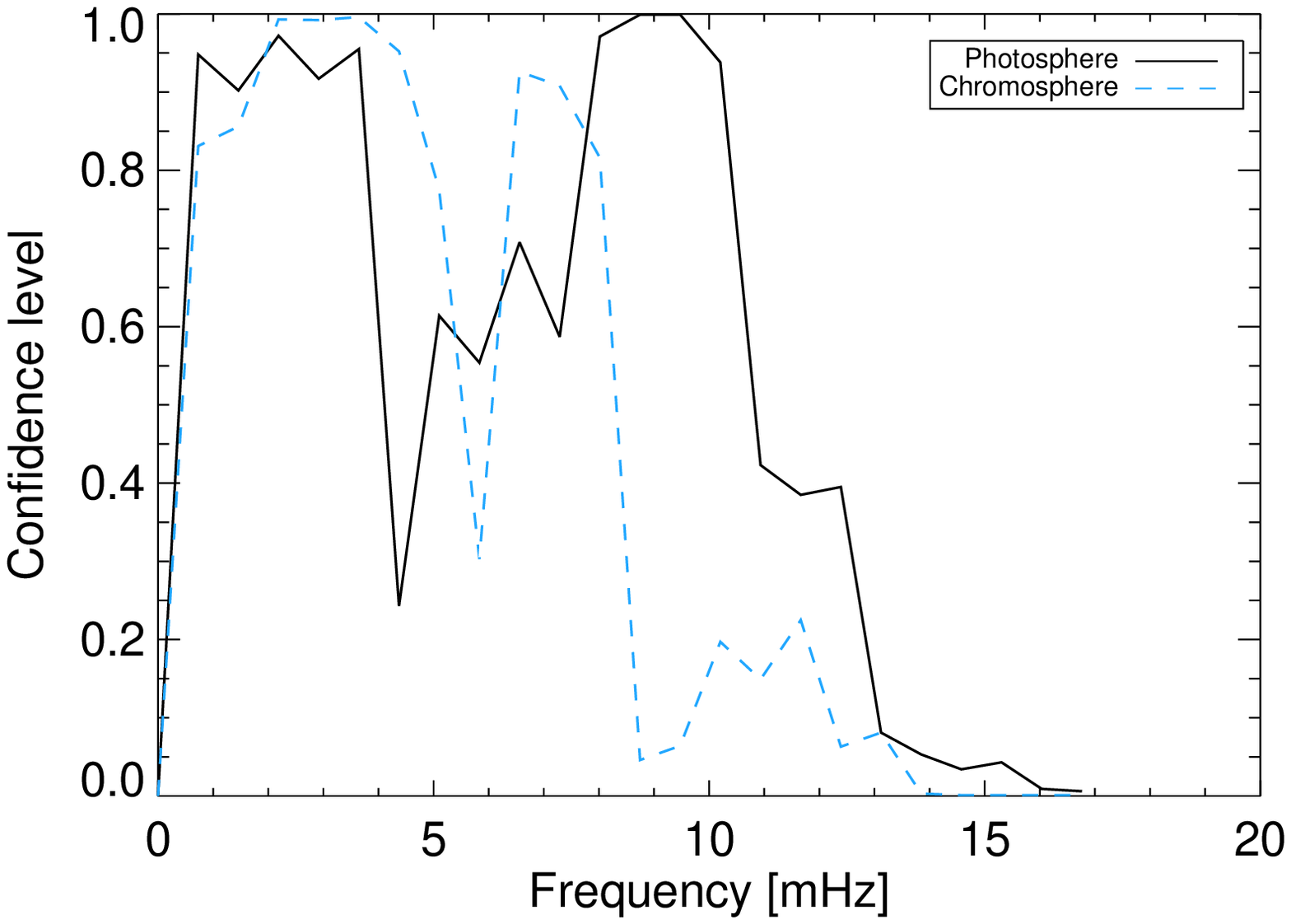}}
 \subfigure [Confidence level magnetic element n.5]{ \includegraphics[width=6cm, clip]{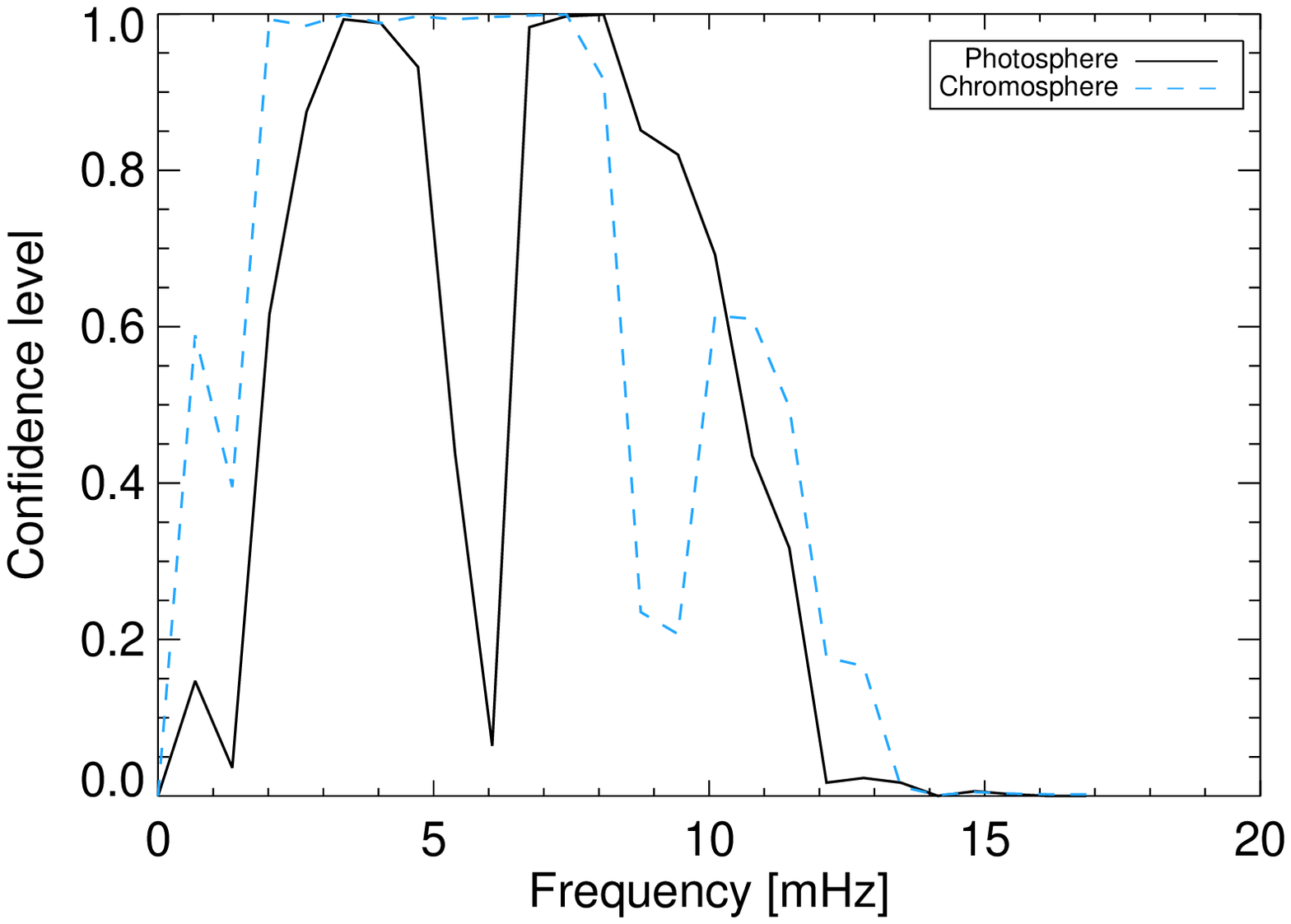}}
 \subfigure [Confidence level magnetic element n.6]{ \includegraphics[width=6cm, clip]{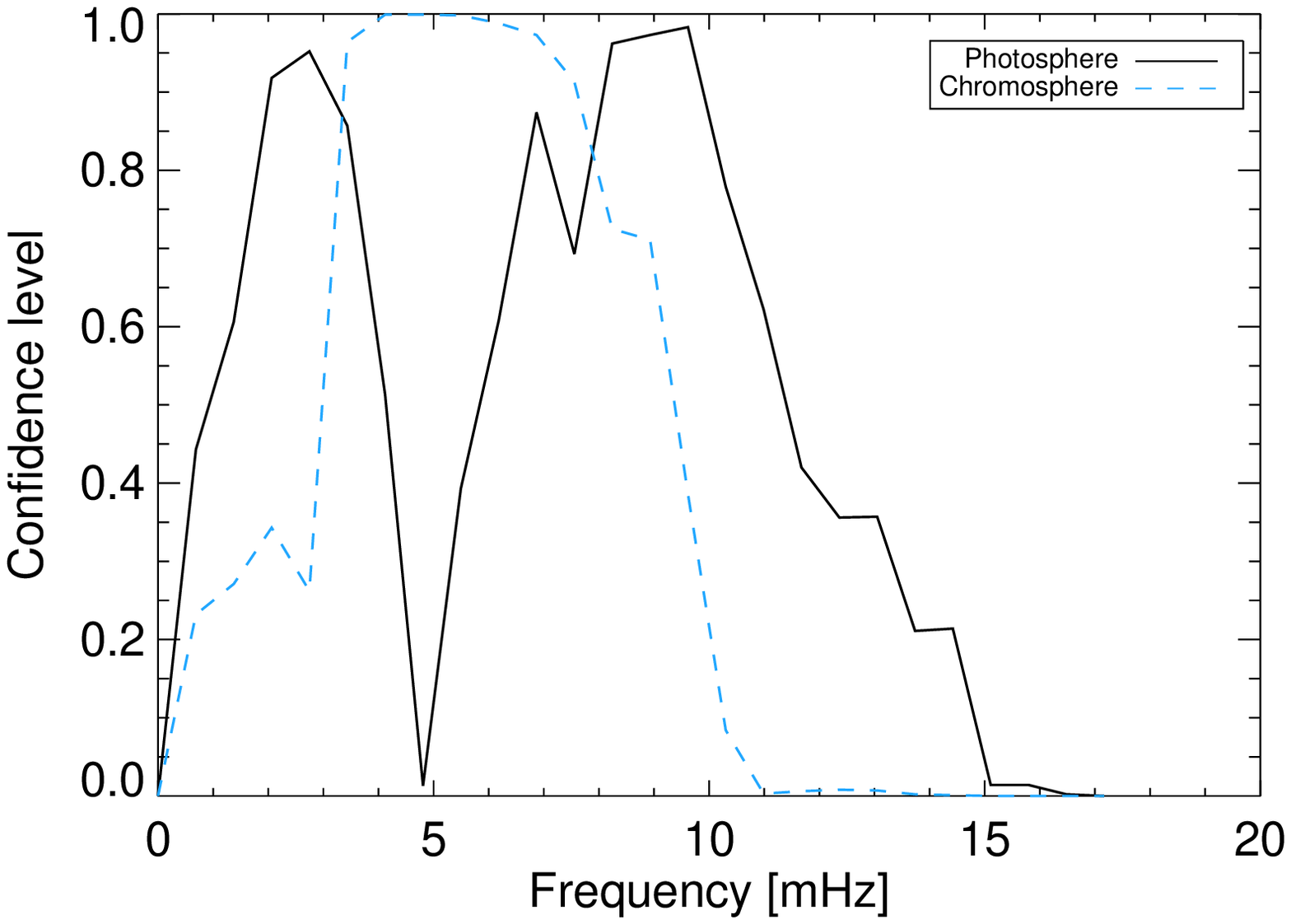} }
   \caption{Photospheric and chromospheric power spectra of horizontal velocity oscillations,  and confidence level plots obtained from a randomization test of the velocity time series.}
    \label{Fig:spectra}
   \end{figure*} 
The photospheric Fe I data were used to estimate the total circular polarization CP within each pixel, which is defined as
\begin{equation}
CP=\int|V(\lambda)|\frac{d\lambda}{I_c},
\label{Eq:total CP}
\end{equation}
where $V(\lambda)$ is the V-Stokes profile, and $I_{c}$ the average continuum intensity.\\
We remark that CP signals can be attributed to the line-of-sight (LOS) magnetic field. 
As a consequence, in what follows small-scale CP signals can be identified with magnetic elements.\\
In Fig. \ref{Fig:maps} we show an example of signed CP polarization map, where the sign is assumed to be that of the blue lobe in Stokes-V profiles, and the corresponding simultaneous broadband image in the core of the Ca II H line. In the same image we also show a photospheric continuum image of the same field taken at $395.3$ nm ($1$ nm width) .\\
Our analysis is based upon the identification and tracking of the longest-lived magnetic elements observed at both the photospheric and chromospheric heights. \\
The studied magnetic elements were identified and tracked by using the well tested YAFTA code \citep{Welsch2003, 2007ApJ...666..576D}, providing it with the CP maps in input. 
The parameters of the code were set to track only those magnetic elements i) whose area was larger than $4\times4$ pixel$^2$ (or equivalently $0".24\times0".24$), which takes into account seeing-induced adaptive optics (AO) residual aberrations that make the FWHM larger than the theoretical (seeing-free) value; ii) whose signal in the CP maps was above two standard deviations of the CP image.
A total of $185$ magnetic elements was found to match these criteria. 
Among them we selected the longest lived (at least $90$ steps or equivalently $\sim 42$ min), and whose chromospheric counterpart was clearly identifiable by visual inspection of the simultaneous data.\\ 
This yielded $35$ magnetic elements, which are marked with yellow circles in Fig. \ref{Fig:maps}.\\
The mean effective radius of the studied magnetic elements, defined as $\sqrt{Area/\pi}$, is $\langle R_e\rangle\simeq280$ km.\\
The magnetic elements selected are, on average, co-aligned within $\sim 2$ pixels. Given the height difference of the sampled atmospheric layers ($\sim 220$ km), this yields an inclination angle of $\sim 22$ degrees. This value is in good agreement with \citet{2014A&A...569A.105J}.
The tracking of the chromospheric features was performed by applying a procedure based on the search for the center of mass of the intensity distribution in windows of area $10\times10$ pixel$^2$ encompassing the intensity enhancement co-spatial to the photospheric feature.\\ 
At each time step, the horizontal velocity of each identified magnetic element was computed as the time derivative of the measured position of the feature.\\
We applied a standard FFT-based analysis to compute the horizontal velocity power spectrum of each photospheric magnetic element and its chromospheric counterpart. 
In doing this, only one component of the horizontal velocity was considered. 
This is to avoid frequency doublings due to the square root operator used in the composition of the two components of the horizontal velocity.
\section{Results}
\label{Sect:Results}
\begin{figure}
  \includegraphics[width=8cm, clip]{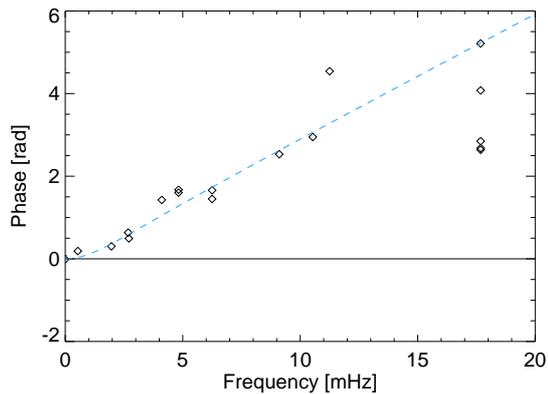}
   \caption{Phase diagram obtained from the tracking of the selected magnetic elements (diamonds). Among the phase spectra used and belonging to the selected elements, only the measurements with a coherence above 0.9 are retained. The dashed line represents the expected phase for propagating kink waves at 6 km/s and with a cutoff frequency of 2.6 mHz, between two layers with a separation of 220 km.}
    \label{Fig:phase}
   \end{figure}

\begin{figure}
  \includegraphics[width=8cm, clip]{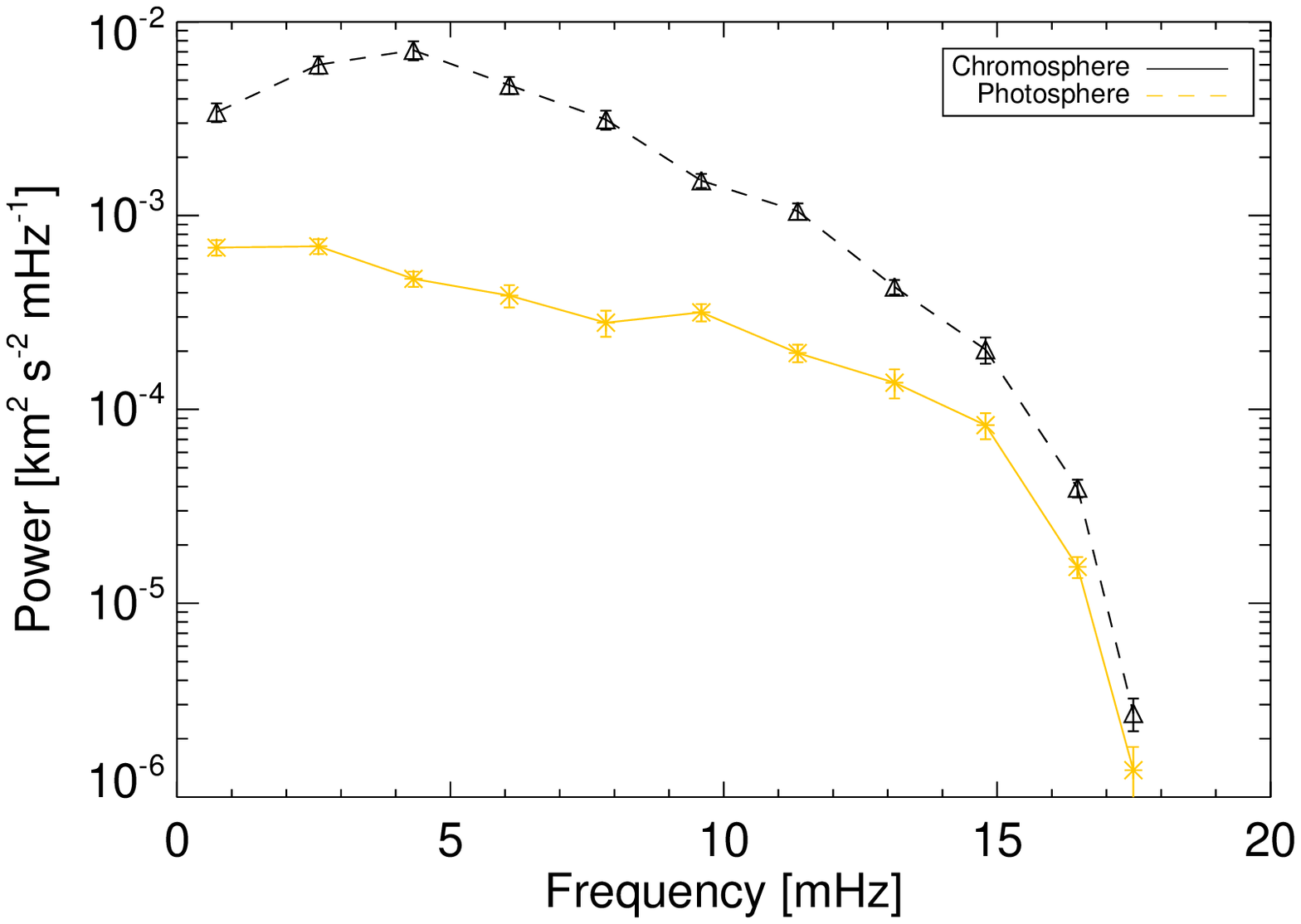}
  \includegraphics[width=8cm, clip]{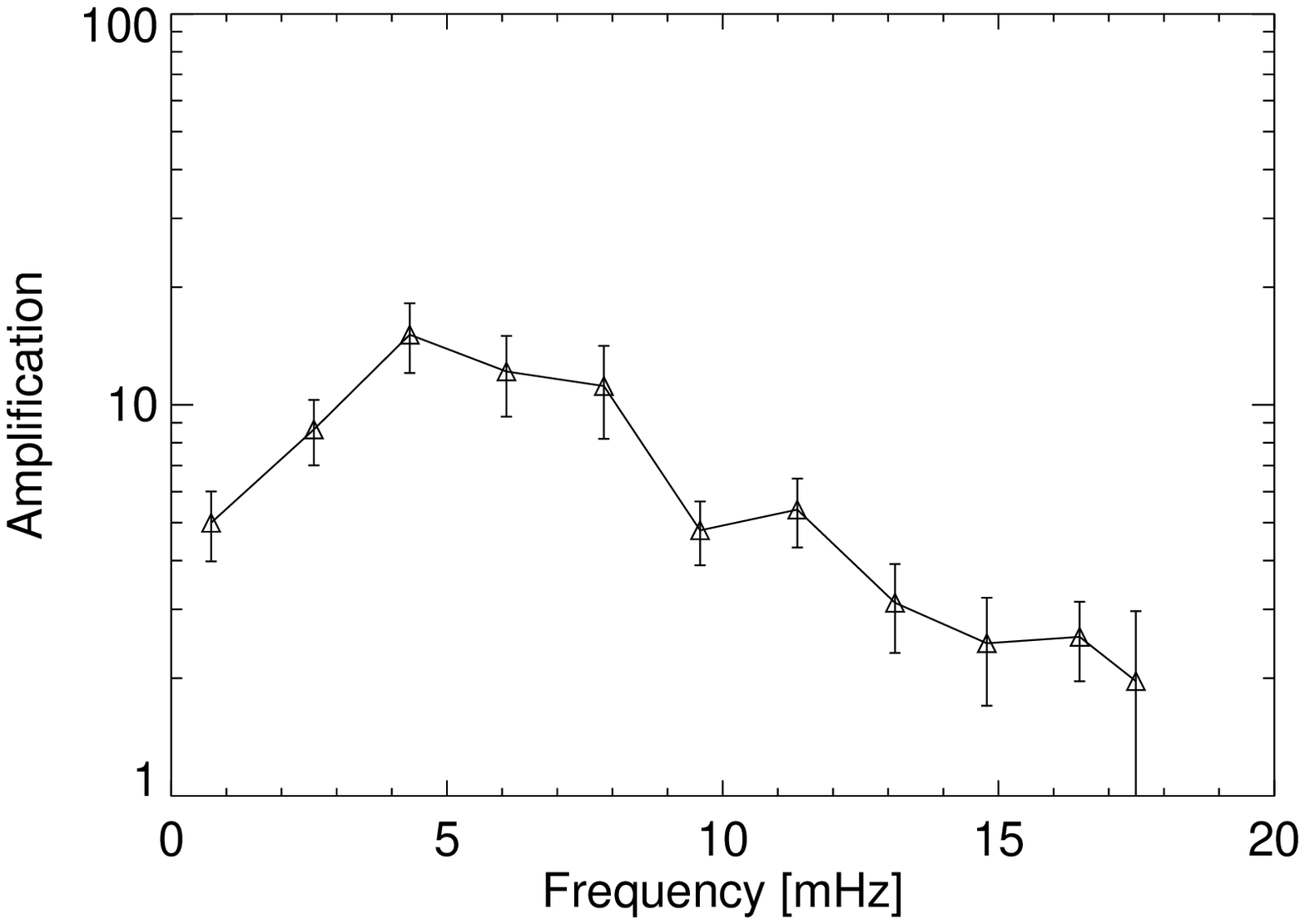}
   \caption{Top: Kink photospheric and chromospheric power spectra obtained from 35 magnetic elements and averaged in bins of 0.8 mHz. 	Bottom: Amplification spectrum.}
  \label{Fig:amplif}
  \end{figure}

In Fig. \ref{Fig:sigmavx} we plot the photospheric (black plus symbols) and chromopsheric (cyan diamonds) RMS (root mean square) velocity amplitude of the magnetic elements as a function of their mean velocity i.e. the mean velocity during the whole lifetime of the element.
As we can see, the velocity amplitudes are appreciably larger in the chromosphere than in the photosphere. This can be also seen in Fig. \ref{Fig:spectra}, where we show the power spectra of the horizontal velocity for six representative cases among the 35 studied. The chromospheric power spectra are largely amplified with respect to their photospheric counterpart. On top of that, they are also characterized by the presence of different peaks from $2-3$ mHz, up to $13-15$ mHz. It is worth noting that the amplification is not constant throughout the whole spectral range. Indeed, different spectral bands are more amplified than others. The different magnetic elements do not share the same peaks, and each of them has its own spectral signature. \\
In Fig. \ref{Fig:spectra}, we also show the confidence levels of each pair of spectra obtained from a randomization test.\\ 
This test consists in the random shuffling of the measurements of the time series associated to the horizontal velocity of each magnetic element. More in detail, this shuffling is repeated a thousand times for each time series and, at each step, the power spectral density is computed.
The number of times a certain peak in the spectrum is larger than that obtained from the original time series (i.e. the correctly ordered one) is counted. This gives the probability that a certain peak in the power spectrum is obtained from a noisy time series, and constitutes the confidence level. \\
As clearly shown by the results obtained, the confidence level associated to most of the prominent peaks in the power spectra have a very high confidence level (above 0.9). For this reason the peaks shown in the power spectra of Fig. \ref{Fig:spectra} can be regarded as highly trusted.\\
We also studied the wave propagation by applying a phase lag analysis. For each magnetic element the FFT phase and coherence spectra between the photospheric and chromospheric heights were estimated. Only those phase measurements in the spectrum whose coherence level is larger than 0.9 are considered for the estimation of the global phase spectrum obtained from the whole sample. This is necessary to avoid the inclusion of unreliable phase measurements in the final global phase spectrum. We remark that a 0.9 coherence level represents a very restrictive threshold, thus ensuring the reliability of our global phase spectrum estimate.\\
The results of this analysis are shown in Fig. \ref{Fig:phase}, where we plot the unwrapped global phase spectrum obtained by collecting the most reliable phase measurements (those with coherence larger than 0.9) in the phase spectra among the 35 selected magnetic elements.\\
In the same figure we also plot for comparison the phase spectrum expected for propagating kink waves in thin flux tubes as obtained from the dispersion relation of Eq. 4 in \citet{PlonerS.R.O.1997}. The phase relation is calculated for a propagation speed of $6$ km/s, a pressure height scale of $\sim 130$ km, and a kink cutoff frequency of $2.6$ mHz. These are in fact typical values expected for small scale flux tubes \citep[see for example][]{1981A&A....98..155S, PlonerS.R.O.1997}.
In our sign convention a positive phase difference represents upward propagation. \\
We then studied the amplification of the power at chromospheric heights. \\
The longest-lived 35 magnetic elements were considered to estimate the average power spectrum in both the photosphere and the chromosphere. In Fig. \ref{Fig:amplif} we show the average chromospheric and photospheric spectra (upper panel) and the amplification spectrum (lower panel). The amplitude of kink waves is largely amplified in the $2-8$ mHz band, with a maximum around $4.2$ mHz where the amplification at the chromospheric heights sampled by the core of Ca II H spectral line reaches a factor of five. The error bars represent the standard error of the mean.\\

\section{Discussions and conclusions}   
In this work we have studied the propagation of kink waves in magnetic elements between the photosphere and the chromosphere. \\
The power spectra of the horizontal velocity of different magnetic elements in the photosphere show different patterns. This is not surprising since the kink mode is largely influenced by the ambient plasma surrounding the magnetic elements, which affects their oscillatory behaviour. This result confirms previous findings by \citet{2013A&A...559A..88S}, obtained from the analysis of Hinode data at lower spatial and temporal resolution.\\ 
However, the results presented here also reveal, for the first time, that the chromospheric spectra show remarkable differences with respect to their photospheric counterparts. The same magnetic element exhibits a completely different pattern of peaks in the photosphere, with respect to the chromospheric power spectrum.\\  
This is an evidence for a non-linear wave propagation regime.\\ 
It is worthwhile noting here that a randomization test demonstrated the reliability of the obtained power spectra with a confidence level larger than $90 \%$. 
We remark that the randomization test also validates the quality of the tracking. More in detail, it is unreasonable to find high confidence levels in the spectra of noisy signals. For this reason, the high confidence levels found in the spectra, together with the large coherence found between the two sampled layers, can be regarded as a good validation of the methods employed for the tracking of the magnetic elements at both heights.\\
We also studied the propagation of the kink waves between the photospheric layer sampled by the Fe I 630.1 nm spectral line and the chromospheric height sampled by the Ca II H spectral data. For this purpose, only the 35 longest-lived magnetic elements were considered in order to maximize the frequency resolution in the power spectra. \\
Moreover, by considering only frequencies in the phase spectra with very high confidence level between the photosphere and the chromosphere, we estimated the global phase spectrum. This is done to ensure the reliability of the phase estimates themselves. However, we note here that, since the propagation of the kink waves is non-linear, in doing this we are only selecting those cases for which a linear regime can be applied. \\
The phase diagram so obtained shows the upward propagation of the kink waves compatible with a velocity of the order of $~6$ km/s, and a cutoff frequency of $~2.6$ mHz. Such value for the cutoff frequency is consistent with the theoretical expectations \citep{1981A&A....98..155S}, and represents a first experimental test of this in the lower solar atmosphere. \\
It is worth remarking here that the inclination of the magnetic elements can significantly affect the theoretical phase relation  shown for comparison in Fig. \ref{Fig:phase}. This is because the dispersion relation of kink waves \citep[see][]{PlonerS.R.O.1997} depends on the acoustic cutoff frequency, which is in turn lowered in inclined magnetic fields \citep[see for example][]{2006ApJ...647L..77M}. In our case, the selected magnetic elements have an inclination angle of $\sim 22$ degrees thus they can be considered nearly vertical. However, the upward propagation of kink waves is unambiguously revealed by the phase analysis of Fig. \ref{Fig:phase}. This result is valid independently of the inclination angle and other physical conditions of the flux tubes that can affect, to some extent, the exact values of the cutoff frequency and the propagation speed \citep{PlonerS.R.O.1997}.
Our findings show that the chromospheric power spectra are largely amplified. The amplification spectrum shows that most of the enhancement is concentrated in the 1-8 mHz band, although, there is no significant power reduction at least up to the limit imposed by the data sampling ($\sim 17$ mHz).\\ 
In the presence of a non-linear propagation regime, kink waves are expected to deposit a significant fraction of energy through dissipation \citep{2014ApJ...784...29M}. The amplification spectrum estimated from the 35 studied magnetic elements demonstrates that this is not the case at the heights sampled by the data analysed in this work (up to $\sim 450$ km above the photosphere).
A dissipation of kink waves would in fact result in a deficit of power at chromospheric heights, with respect to photospheric ones. For this reason, we argue that, in absence of deficits of power in the chromospheric spectra, no dissipation should be expected up to approximatively $450$ km above the photosphere, where the core of the Ca II H spectral line forms. \\
\citet{2014ApJ...784...29M} have observed the dissipation of kink waves between chromosphere and the corona. In this context, the results presented here demonstrate that the energy contained in the MHD kink waves in small magnetic elements is not dissipated in the low atmosphere but flows entirely to heights which are above the altitude of formation of the core of the Ca II H spectral line. This may have implications for the energy balance of both chromospheric and, in particular, coronal heights, to where most of the energy flows. Finally, we note that the lack of damping is also evidence for the fact that the kink waves do not interact with other modes during the propagation between the heights sampled by our observations.

\begin{acknowledgements}
This work has been partly supported by the "Progetti di ricerca INAF di Rilevante Interesse Nazionale" (PRIN-INAF) 2010 and PRIN-MIUR 2012 (prot. 2012P2HRCR) entitled "Il sole attivo ed i suoi effetti sul clima dello spazio e della terra" grants, funded by the Italian National Institute for Astrophysics (INAF) and Ministry of Education, Universities and Research (MIUR), respectively. This study received funding from the European Unions Seventh Programme for Research, Technological Development and Demonstration, under the Grant Agreements of the eHEROES (n 284461, www.eheroes.eu), SOLARNET (n 312495, www.solarnet-east.eu). We also thank Luc Rouppe van der Voort and Alice Cristaldi for their help in supporting the data reduction, and Serena Criscuoli for the support during the data acquisition. M. S. and F. G. are grateful to Robertus Erdélyi for the useful discussions and suggestions provided, and to Ilaria Ermolli for a critical reading of the manuscript.  \\
\end{acknowledgements}

\bibliographystyle{aa}
\bibliography{lib}

\begin{thebibliography}{47}
\expandafter\ifx\csname natexlab\endcsname\relax\def\natexlab#1{#1}\fi

\bibitem[{{Abramenko} {et~al.}(2011){Abramenko}, {Carbone}, {Yurchyshyn},
  {Goode}, {Stein}, {Lepreti}, {Capparelli}, \&
  {Vecchio}}]{2011ApJ...743..133A}
{Abramenko}, V.~I., {Carbone}, V., {Yurchyshyn}, V., {et~al.} 2011, \apj, 743,
  133

\bibitem[{{Bonet} {et~al.}(2012){Bonet}, {Cabello}, \& {S{\'a}nchez
  Almeida}}]{2012A&A...539A...6B}
{Bonet}, J.~A., {Cabello}, I., \& {S{\'a}nchez Almeida}, J. 2012, \aap, 539, A6

\bibitem[{Choudhuri {et~al.}(1993)Choudhuri, Dikpati, \&
  Banerjee}]{1993ApJ...413..811C}
Choudhuri, A.~R., Dikpati, M., \& Banerjee, D. 1993, \apj, 413, 811

\bibitem[{{de la Cruz Rodr{\'{\i}}guez} {et~al.}(2015){de la Cruz
  Rodr{\'{\i}}guez}, {L{\"o}fdahl}, {S{\"u}tterlin}, {Hillberg}, \& {Rouppe van
  der Voort}}]{2015A&A...573A..40D}
{de la Cruz Rodr{\'{\i}}guez}, J., {L{\"o}fdahl}, M.~G., {S{\"u}tterlin}, P.,
  {Hillberg}, T., \& {Rouppe van der Voort}, L. 2015, \aap, 573, A40

\bibitem[{{DeForest} {et~al.}(2007){DeForest}, {Hagenaar}, {Lamb}, {Parnell},
  \& {Welsch}}]{2007ApJ...666..576D}
{DeForest}, C.~E., {Hagenaar}, H.~J., {Lamb}, D.~A., {Parnell}, C.~E., \&
  {Welsch}, B.~T. 2007, \apj, 666, 576

\bibitem[{Edwin \& Roberts(1983)}]{Edwin1983}
Edwin, P. \& Roberts, B. 1983, Sol. Phys., 88, 179

\bibitem[{{Erd{\'e}lyi} \& {Fedun}(2007)}]{2007Sci...318.1572E}
{Erd{\'e}lyi}, R. \& {Fedun}, V. 2007, Science, 318, 1572

\bibitem[{{Faurobert} {et~al.}(2012){Faurobert}, {Ricort}, \&
  {Aime}}]{2012A&A...548A..80F}
{Faurobert}, M., {Ricort}, G., \& {Aime}, C. 2012, \aap, 548, A80

\bibitem[{Fedun {et~al.}(2011)Fedun, Shelyag, \& Erd\'{e}lyi}]{Fedun2011}
Fedun, V., Shelyag, S., \& Erd\'{e}lyi, R. 2011, ApJ, 727, 17

\bibitem[{{Giannattasio} {et~al.}(2014{\natexlab{a}}){Giannattasio},
  {Berrilli}, {Biferale}, {Del Moro}, {Sbragaglia}, {Bellot Rubio}, {Go{\v
  s}i{\'c}}, \& {Orozco Su{\'a}rez}}]{fabio04}
{Giannattasio}, F., {Berrilli}, F., {Biferale}, L., {et~al.}
  2014{\natexlab{a}}, \aap, 569, A121

\bibitem[{{Giannattasio} {et~al.}(2013){Giannattasio}, {Del Moro}, {Berrilli},
  {Bellot Rubio}, {Gos{\#728}i{\'c}}, \& {Orozco
  Su{\'a}rez}}]{2013ApJ...770L..36G}
{Giannattasio}, F., {Del Moro}, D., {Berrilli}, F., {et~al.} 2013, \apjl, 770,
  L36

\bibitem[{{Giannattasio} {et~al.}(2014{\natexlab{b}}){Giannattasio},
  {Stangalini}, {Berrilli}, {Del Moro}, \& {Bellot
  Rubio}}]{2014arXiv1405.0677G}
{Giannattasio}, F., {Stangalini}, M., {Berrilli}, F., {Del Moro}, D., \&
  {Bellot Rubio}, L. 2014{\natexlab{b}}, \apj, 788, 137

\bibitem[{Hasan {et~al.}(2003)Hasan, Kalkofen, van Ballegooijen, \&
  Ulmschneider}]{Hasan2003}
Hasan, S.~S., Kalkofen, W., van Ballegooijen, A.~A., \& Ulmschneider, P. 2003,
  ApJ, 585, 1138

\bibitem[{{Hasan} \& {van Ballegooijen}(2008)}]{2008ApJ...680.1542H}
{Hasan}, S.~S. \& {van Ballegooijen}, A.~A. 2008, ApJ, 680, 1542

\bibitem[{{Jafarzadeh} {et~al.}(2013){Jafarzadeh}, {Solanki}, {Feller}, {Lagg},
  {Pietarila}, {Danilovic}, {Riethm{\"u}ller}, \& {Mart{\'{\i}}nez
  Pillet}}]{2013A&A...549A.116J}
{Jafarzadeh}, S., {Solanki}, S.~K., {Feller}, A., {et~al.} 2013, \aap, 549,
  A116

\bibitem[{{Jafarzadeh} {et~al.}(2014){Jafarzadeh}, {Solanki}, {Lagg}, {Bellot
  Rubio}, {van Noort}, {Feller}, \& {Danilovic}}]{2014A&A...569A.105J}
{Jafarzadeh}, S., {Solanki}, S.~K., {Lagg}, A., {et~al.} 2014, \aap, 569, A105

\bibitem[{{Jess} {et~al.}(2012){Jess}, {Shelyag}, {Mathioudakis}, {Keys},
  {Christian}, \& {Keenan}}]{2012ApJ...746..183J}
{Jess}, D.~B., {Shelyag}, S., {Mathioudakis}, M., {et~al.} 2012, ApJ, 746, 183

\bibitem[{{Kalkofen}(1997)}]{1997ApJ...486L.145K}
{Kalkofen}, W. 1997, \apjl, 486, L145

\bibitem[{{Keys} {et~al.}(2011){Keys}, {Mathioudakis}, {Jess}, {Shelyag},
  {Crockett}, {Christian}, \& {Keenan}}]{2011ApJ...740L..40K}
{Keys}, P.~H., {Mathioudakis}, M., {Jess}, D.~B., {et~al.} 2011, \apjl, 740,
  L40

\bibitem[{Khomenko {et~al.}(2008)Khomenko, Collados, \& Felipe}]{Khomenko2008}
Khomenko, E., Collados, M., \& Felipe, T. 2008, Sol. Phys., 251, 589

\bibitem[{{Lagg} {et~al.}(2010){Lagg}, {Solanki}, {Riethm{\"u}ller},
  {Mart{\'{\i}}nez Pillet}, {Sch{\"u}ssler}, {Hirzberger}, {Feller}, {Borrero},
  {Schmidt}, {del Toro Iniesta}, {Bonet}, {Barthol}, {Berkefeld}, {Domingo},
  {Gandorfer}, {Kn{\"o}lker}, \& {Title}}]{2010ApJ...723L.164L}
{Lagg}, A., {Solanki}, S.~K., {Riethm{\"u}ller}, T.~L., {et~al.} 2010, \apjl,
  723, L164

\bibitem[{{Lepreti} {et~al.}(2012){Lepreti}, {Carbone}, {Abramenko},
  {Yurchyshyn}, {Goode}, {Capparelli}, \& {Vecchio}}]{2012ApJ...759L..17L}
{Lepreti}, F., {Carbone}, V., {Abramenko}, V.~I., {et~al.} 2012, \apjl, 759,
  L17

\bibitem[{{McIntosh} {et~al.}(2011){McIntosh}, {de Pontieu}, {Carlsson},
  {Hansteen}, {Boerner}, \& {Goossens}}]{2011Natur.475..477M}
{McIntosh}, S.~W., {de Pontieu}, B., {Carlsson}, M., {et~al.} 2011, \nat, 475,
  477

\bibitem[{McIntosh \& Jefferies(2006)}]{2006ApJ...647L..77M}
McIntosh, S.~W. \& Jefferies, S.~M. 2006, \apjl, 647, L77

\bibitem[{{Morton} {et~al.}(2013){Morton}, {Verth}, {Fedun}, {Shelyag}, \&
  {Erd{\'e}lyi}}]{2013ApJ...768...17M}
{Morton}, R.~J., {Verth}, G., {Fedun}, V., {Shelyag}, S., \& {Erd{\'e}lyi}, R.
  2013, \apj, 768, 17

\bibitem[{{Morton} {et~al.}(2014){Morton}, {Verth}, {Hillier}, \&
  {Erd{\'e}lyi}}]{2014ApJ...784...29M}
{Morton}, R.~J., {Verth}, G., {Hillier}, A., \& {Erd{\'e}lyi}, R. 2014, \apj,
  784, 29

\bibitem[{Musielak {et~al.}(1989)Musielak, Rosner, \&
  Ulmschneider}]{Musielak1989}
Musielak, Z.~E., Rosner, R., \& Ulmschneider, P. 1989, ApJ, 337, 470

\bibitem[{Musielak \& Ulmschneider(2003{\natexlab{a}})}]{Musielak2003a}
Musielak, Z.~E. \& Ulmschneider, P. 2003{\natexlab{a}}, A\&A, 400, 1057

\bibitem[{Musielak \& Ulmschneider(2003{\natexlab{b}})}]{Musielak2003}
Musielak, Z.~E. \& Ulmschneider, P. 2003{\natexlab{b}}, A\&A, 406, 725

\bibitem[{{Nutto} {et~al.}(2012){Nutto}, {Steiner}, {Schaffenberger}, \&
  {Roth}}]{2012A&A...538A..79N}
{Nutto}, C., {Steiner}, O., {Schaffenberger}, W., \& {Roth}, M. 2012, \aap,
  538, A79

\bibitem[{{Ploner, S. R. O.} \& {Solanki, S. K.}(1997)}]{PlonerS.R.O.1997}
{Ploner, S. R. O.} \& {Solanki, S. K.} 1997, Astronomy and Astrophysics

\bibitem[{Roberts(1983)}]{Roberts1983}
Roberts, B. 1983, Sol. Phys., 87, 77

\bibitem[{{Roberts} \& {Webb}(1978)}]{1978SoPh...56....5R}
{Roberts}, B. \& {Webb}, A.~R. 1978, \solphys, 56, 5

\bibitem[{{Sander} \& {Yorke}(2009)}]{2009arXiv0910.3570S}
{Sander}, E. \& {Yorke}, J.~A. 2009, ArXiv e-prints 0910.3570

\bibitem[{{Sander} \& {Yorke}(2010)}]{2010arXiv1002.3363S}
{Sander}, E. \& {Yorke}, J.~A. 2010, ArXiv e-prints 1002.3363

\bibitem[{{Scharmer}(2006)}]{2006A&A...447.1111S}
{Scharmer}, G.~B. 2006, \aap, 447, 1111

\bibitem[{{Scharmer} {et~al.}(2003){Scharmer}, {Dettori}, {Lofdahl}, \&
  {Shand}}]{2003SPIE.4853..370S}
{Scharmer}, G.~B., {Dettori}, P.~M., {Lofdahl}, M.~G., \& {Shand}, M. 2003, in
  Society of Photo-Optical Instrumentation Engineers (SPIE) Conference Series,
  Vol. 4853, Innovative Telescopes and Instrumentation for Solar Astrophysics,
  ed. S.~L. {Keil} \& S.~V. {Avakyan}, 370--380

\bibitem[{{Scharmer} {et~al.}(2008){Scharmer}, {Narayan}, {Hillberg}, {de la
  Cruz Rodr{\'{\i}}guez}, {L{\"o}fdahl}, {Kiselman}, {S{\"u}tterlin}, {van
  Noort}, \& {Lagg}}]{2008ApJ...689L..69S}
{Scharmer}, G.~B., {Narayan}, G., {Hillberg}, T., {et~al.} 2008, \apjl, 689,
  L69

\bibitem[{{Spruit}(1981)}]{1981A&A....98..155S}
{Spruit}, H.~C. 1981, \aap, 98, 155

\bibitem[{{Stangalini}(2014)}]{2014A&A...561L...6S}
{Stangalini}, M. 2014, \aap, 561, L6

\bibitem[{{Stangalini} {et~al.}(2013{\natexlab{a}}){Stangalini}, {Berrilli}, \&
  {Consolini}}]{2013A&A...559A..88S}
{Stangalini}, M., {Berrilli}, F., \& {Consolini}, G. 2013{\natexlab{a}}, \aap,
  559, A88

\bibitem[{{Stangalini} {et~al.}(2014){Stangalini}, {Consolini}, {Berrilli}, {De
  Michelis}, \& {Tozzi}}]{refId0}
{Stangalini}, M., {Consolini}, G., {Berrilli}, F., {De Michelis}, P., \&
  {Tozzi}, R. 2014, \aap, 569, A102

\bibitem[{{Stangalini} {et~al.}(2013{\natexlab{b}}){Stangalini}, {Solanki},
  {Cameron}, \& {Mart{\'{\i}}nez Pillet}}]{2013A&A...554A.115S}
{Stangalini}, M., {Solanki}, S.~K., {Cameron}, R., \& {Mart{\'{\i}}nez Pillet},
  V. 2013{\natexlab{b}}, \aap, 554, A115

\bibitem[{Steiner {et~al.}(1998)Steiner, Grossmann-Doerth, Knoelker, \&
  Sch{\"u}ssler}]{1998ApJ...495..468S}
Steiner, O., Grossmann-Doerth, U., Knoelker, M., \& Sch{\"u}ssler, M. 1998,
  ApJ, 495, 468

\bibitem[{van Noort {et~al.}(2005)van Noort, {Rouppe van der Voort}, \&
  L\"{o}fdahl}]{MSnoort05}
van Noort, M., {Rouppe van der Voort}, L., \& L\"{o}fdahl, M.~G. 2005,
  $\backslash$solphys, 228, 191

\bibitem[{{Vigeesh} {et~al.}(2012){Vigeesh}, {Fedun}, {Hasan}, \&
  {Erd{\'e}lyi}}]{2012ApJ...755...18V}
{Vigeesh}, G., {Fedun}, V., {Hasan}, S.~S., \& {Erd{\'e}lyi}, R. 2012, \apj,
  755, 18

\bibitem[{Welsch \& Longcope(2003)}]{Welsch2003}
Welsch, B.~T. \& Longcope, D.~W. 2003, ApJ, 588, 620

\end{thebibliography}
\end{document}